\documentclass[letterpaper, journal, twoside, 12pt]{support/IEEEtran}

\pdfoutput=1
\usepackage{mathrsfs}
\usepackage{graphics}
\usepackage{mathptmx}
\usepackage{times}
\usepackage{epstopdf}
\usepackage{color}
\usepackage{enumerate}
\usepackage{float}
\usepackage{indentfirst}
\usepackage{multirow}
\usepackage[noadjust]{cite}
\usepackage{amsthm,amsfonts}
\usepackage{psfrag}
\usepackage{epsfig}
\usepackage{graphicx,subfigure}
\usepackage{multirow}
\usepackage{setspace}
\usepackage{caption}
\usepackage{booktabs} 
\usepackage[fleqn]{amsmath}
\usepackage{amssymb}
\usepackage{algorithm}
\usepackage[noend]{algpseudocode}
\usepackage{dsfont}

\usepackage{url}
\usepackage{tikz}
\usepackage{subeqnarray}
\usepackage{cases}
\usepackage{color}

\usepackage[shortlabels]{enumitem}
\usepackage{exscale}
\usepackage{relsize}
\usepackage{bm}
\usepackage{enumerate}
\usepackage{enumitem}

\normalbaselineskip=\baselineskip
\renewcommand{\baselinestretch}{1.8}
\newtheorem{myTheo}{Theorem}

\newtheorem{remark}[myTheo]{Remark}

\graphicspath{{figures/}}
\DeclareGraphicsExtensions{.pdf,.png,.jpg,.eps}
\IEEEoverridecommandlockouts

\title{\LARGE \bf Towards Safe and Efficient Swarm-Human Collaboration: A Hierarchical Multi-Agent Pickup and Delivery framework}

\author{
  \vskip 1em
  {
  Xin Gong, \emph{Graduate Student Member, IEEE},
  Tieniu Wang, \emph{}
    Yukang Cui, \emph{Member, IEEE},
  and Tingwen Huang, \emph{Fellow,~IEEE}

  }

  \thanks{
    This work is partially supported by the National Natural Science Foundation of China under Grant 61903258, 61973156, 61603180, Qatar National Research Fund NPRP12C-0814-190012. 

X. Gong is with the Department of Mechanical Engineering, The University of Hong Kong, Pokfulam Road, Hong Kong (e-mail: {\tt\small gongxin@connect.hku.hk}).

T. Wang and Y. Cui are with the College of Mechatronics and Control Engineering, Shenzhen University, Shenzhen, 518060, China (e-mail: {\tt\small szuwtn,cuiyukang@gmail.com}).

T. Huang is with Texas A\&M University at Qatar, Doha, 23874, Qatar (e-mail: {\tt\small tingwen.huang@qatar.tamu.edu}).

  }
}

\begin{document}
    \maketitle

    \begin{abstract}
The multi-Agent Pickup and Delivery (MAPD) problem is crucial in the realm of Intelligent Storage Systems (ISSs), where multiple robots are assigned with time-varying, heterogeneous, and potentially uncertain tasks. When it comes to Human-Swarm Hybrid System ((HS)$_2$), robots and human workers will accomplish the MAPD tasks in collaboration. Herein, we propose a  Human-Swarm Hybrid System Pickup and Delivery ((HS)$_2$PD) framework, which is predominant in future ISSs. A two-layer decision framework based on the prediction horizon window is established in light of the unpredictability of human behavior and the dynamic changes of tasks. The first layer is a two-level programming problem to solve the problems of mode assignment and TA. The second layer is devoted to the exact path of each agent via solving mixed-integer programming (MIP) problems. An integrated algorithm for the (HS)$_2$PD problem is summarized. The practicality and validity of the above algorithm are illustrated via a numerical simulation example towards (HS)$_2$PD tasks.
    \end{abstract}
\begin{IEEEkeywords}
 Pickup and Delivery, Human-Swarm Hybrid System, Mixed Integer Programming, Task allocation, Path finding
\end{IEEEkeywords}
\section{Introduction}


\IEEEPARstart{T}{he} last decade has witnessed the rapid developments of Intelligent Storage Systems (ISSs) \cite{ifr2017executive,probst2015service, chen2015paired, chen2014model, ma2017lifelong, chen2021integrated, liu2019task, yamauchi2021path, ma2019lifelong}, which brings forward new demands and challenges to safe, human-friendly, and environment-adaptable logistics robot swarms.  {\color{blue} Also, the solution of logistics robots has become more and more unaffordable, which makes it competitive for small/medium-sized companies to be employed.} Hence, Human-Swarm Hybrid System ((HS)$_2$), where a swarm of robots working together with human forces, are becoming the new frontier in ISS \cite{ifr2017executive,probst2015service}.  {\color{blue}The (HS)$_2$ solution, powered by the advanced communication, computation ,and control (3C) technologies, is one of the key components for Industry 4.0 \cite{kagermann2013recommendations}.} With the assistance of (HS)$_2$, the ISS will be more efficient, economical, and user-friendly through integrated automation. Thus, it is a must to investigate and develop (HS)$_2$, in order to improve the swarm's flexibility and inter-operability w.r.t. human workers. However, two inevitable problems arose when developing collective robot swarms:
\begin{enumerate}
  \item \textbf{Safety of human workers}: First, safety issues account for the main primary challenges when implementing collaboration between human workers and robots, which cannot be neglected. Indeed, there exist few eliminating fences between robots and human workers. In order to ensure that the whole ISS is operating safely, these robots should actively avoid any possible collisions among  (HS)$_2$, especially collisions between human workers and robots.
  \item \textbf{Counterproductive behavior of human workers}:  One of the most significant advantages brought by collaborative robot swarms lies in the opportunity to deal with the intentional/unintentional counterproductive behavior of human workers. {\color{blue}In the framework of  (HS)$_2$, the productivity of workers can be enhanced, their stress and fatigue can be reduced, and their missing work can be remedied with the assistance of swarms.} 
\end{enumerate}

\begin{figure}[thpb]  
	\centering 
	\subfigure[Transport workers]{\label{fig01}\includegraphics[scale=0.24]{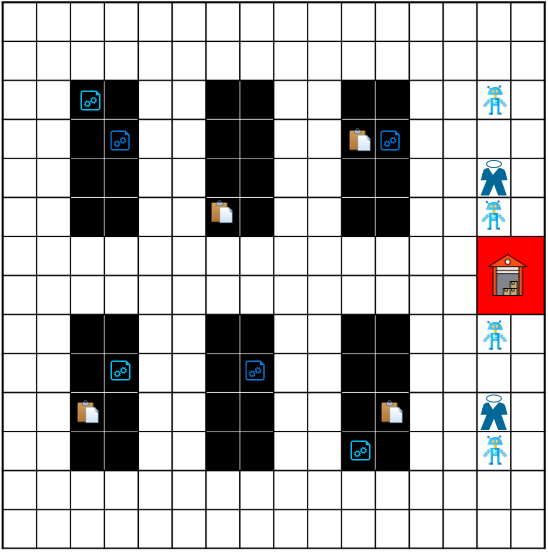} }~~~~~
	\subfigure[Transport robots]{\label{fig02}\includegraphics[scale=0.30]{change.png}} 
 	\caption{The PD task accomplished via different means \cite{da2021robotic}.}
	\label{fig0} 
\end{figure} 

Herein, we aim to design practical algorithms for the robot swarm in the (HS)$_2$ to improve efficiency and resilience.

\subsection{Related Works}

In recent years, the logistics industry has developed rapidly. In many environments such as warehouses, docks, communities, and office buildings, there exists a large number of parcels to be transported. Such problems can be summarized as pickup and delivery (PD) problems  \cite{savelsbergh1995general}. {\color{blue}The PD problems are usually abstracted as vehicle routing problems}, in which parcels should be delivered between a start point and an endpoint \cite{berbeglia2010dynamic}. There are many classifications of PD problems \cite{parragh2006survey}, such as synchronous pickup and delivery \cite{chen2015paired}, asynchronous pickup and delivery \cite{chen2014model}, simultaneous multiple pickup and single delivery \cite{berbeglia2010dynamic}, and simultaneous multiple pickup and multiple delivery \cite{chen2015paired}. {\color{blue}A variant of PD named Multi-Agent Pickup and Delivery (MAPD) problem has been extensively studied in recent works  \cite{ma2017lifelong, chen2021integrated, liu2019task, yamauchi2021path, ma2019lifelong} with distributed robot swarms involved.}

However, most of the research is focused on a fixed task pattern. Very little work has been done to solve the problem of mixing PD in different modes, which is likely to be needed in actual production \cite{dragomir2018multidepot}. In \cite{dragomir2018multidepot}, it is pointed out that a switch of transportation modes is required under the multimodality feature of realistic transportation networks.
On the other hand, the PD problems in these studies are mostly single-agent or homogeneous multi-agents. However, the development of human-swarm interaction research \cite{roundtree2019transparency} brings new challenges to the PD issues. To our best knowledge, no one has taken human-robot collaboration and PD problems into account simultaneously. However, it is common for human workers and robots to work together in the real logistics and freight field. This paper aims to find a safe and efficient solution to the complete (HS)$_2$ Pickup and Delivery ($\rm{(HS)_2PD}$) problem. Fig. \ref{fig0} shows a form of existence of robots and human workers in the \textbf{Problem (HS)$_2$PD}.

The MAPD problem has been proven to be an NP-hard problem \cite{hochba1997approximation}. Fortunately, a vast number of studies \cite{chen2014model, ma2017lifelong, liu2019task, yamauchi2021path, ma2019lifelong} have proposed solutions, which decouple the MAPD problem into the multi-agent task allocation {\color{blue}(TA)} problem \cite{korsah2013comprehensive} and the multi-agent path finding {\color{blue}(PF)} problem \cite{stern2019multi}. But at the same time, {\color{blue}some many new requirements and factors need to be considered \cite{jun2021pickup, track2020research}.} In \cite{jeon2017multi}, a multi-robot TA problem with multi-task load capability in a hospital environment is discussed. It focuses on improving the cooperative efficiency of robot swarms without considering the dynamic changes of tasks. In contrast, the authors in \cite{fransen2020dynamic} propose a dynamic method, but only to solve the deadlock problem of PF. In order to deal with dynamic changes in both TA and PF, a two-layer structure control system for post-disaster inspection of UAVs is proposed in \cite{fu2021real}. A human-machine collaborative assembly planning is processed in layers in \cite{johannsmeier2016hierarchical}. However, this work is geared towards the assembly process. In \cite{schouwenaars2004receding}, the path is planned based on a horizontal backoff strategy, which considers dynamic environments. {\color{blue}Considering the practical problem in more detail,} the vehicle PF problem with time windows is solved using mixed-integer programming (MIP) in \cite{zuo2018linear}. {\color{blue} Few articles take into account the human factor for MAPD problem, so we focus on the  \textbf{Problem (HS)$_2$PD}.}

\subsection{Main contributions}
Therefore,  we propose a safe and efficient layered human-machine collaborative pickup and delivery framework for the \textbf{Problem (HS)$_2$PD}, whose characteristics and  contributions can be summarized as follows:

\begin{enumerate}
  \item \textbf{Hierarchical structure}: This research decouples the PD problem into MA and PF problems and then proposes a two-layer optimization scheme based on mixed-integer programming. Secondly, according to the characteristics of \textbf{Problem (HS)$_2$PD}, the task is divided into several modes. As shown in Table \ref{tab1}, the first layer is divided into high and low layers, which respectively solve the problem of mode assignment and TA. The second layer is a mixed-integer programming problem that solves the exact PF problem.
  \item \textbf{(HS)$_2$}: In the modeling of mixed-integer programming problems, tasks are divided between robots and human workers simultaneously. We take into account human fatigue and inertia at work. As a result, a more humanized assignment result can be produced, and human productivity can be improved.
  \item \textbf{{\color{blue}Dynamic response}}: This paper adopts a rolling optimization method based on the prediction horizon. {\color{blue}The prediction horizon strategy sets the prediction horizon and the update time step}, which could respond to the unpredictable task changes. Notice that the optimality of the current decision is guaranteed, which could remedy the mistakes made by human workers.
  \item \textbf{{\color{blue}Safety issue}}: {\color{blue}The framework proposed in this paper fully considers the collision avoidance constraints between agents. The result of PF for the robot swarm is a set of conflict-free paths, reducing the risk of collisions between robots. A robot will make an emergency stop and re-plan the route through a dynamic update mechanism when it detects that there is a human worker on the road ahead. There will be no collision between the robots and human workers. The safety of (HS)$_2$ is guaranteed.}
\end{enumerate}

This article is organized as follows. In section \ref{section2}, the \textbf{Problem (HS)$_2$PD} is formulated, and three modes for the (HS)$_2$ is explained. In section \ref{sec3}, we propose a two-layer control structure {\color{blue}to solve the \textbf{Problem (HS)$_2$PD}.} Two optimization frameworks are developed for the first and second levels, namely the mixed-integer programming model for TA and the mixed-integer programming model for PF. In Section \ref{SecSm}, a MATLAB simulation is conducted {\color{blue} to verify the effectiveness of the method proposed in the article}. Section \ref{sec5} concludes the article.

\noindent\textbf{Notations:}
$\mathcal{A}\cup \mathcal{B}$,$\mathcal{A}\cap \mathcal{B}$, $\mathcal{A}\backslash \mathcal{B}$ are union set, joint set, and difference set for any two given sets $\mathcal{A}$ and $\mathcal{B}$, respectively. Denote the index set of sequential integers as {\color{blue}$\textbf{I}[m,n]=\{m,m+1,\ldots,n\}$} where $m<n$ are two natural numbers. $\mathcal{S}=\{s_i\}_{i \in \textbf{I}[1,n]}$ defines a set of $n$ elements, and $s_i$ represents one of the elements in the set $\mathcal{S}$.

\label{introduction}

\section{Problem Formulation}\label{section2}

{\color{blue}In this section, we mathematically model the \textbf{Problem (HS)$_2$PD} and describe the requirements and properties of the \textbf{Problem (HS)$_2$PD} in detail.}

Consider the \textbf{Problem (HS)$_2$PD} that multiple robots and human workers collectively transport a set of goods with different time requirements,  from their initial positions to associated destinations. {\color{blue}Each agent has a limited carrying capacity, moves with a constant speed for transporting the tasks while avoiding collisions, stops moving after completing its tasks, and waits to accept the next set of tasks.} The difference from the MAPD problem in \cite{ma2017lifelong} is mainly reflected in two aspects.

One aspect is that according to the nature of the task, the task can be divided into two modes, namely \emph{pickup task mode} and \emph{delivery task mode}, respectively.
Each workstation may publish one of two modes of tasks simultaneously. The difference between the two modes will be explained in subsection \ref{setting}.
Another aspect is that the working capabilities of human workers and robots are different. {\color{blue}The differences in the working characteristics of human workers and robots can affect the results of task allocation, which will be analyzed in detail in subsection \ref{norm_TA}.} {\color{blue}Although the task request of a workstation can be responded to by any one of human workers and robots without special requirements,} due to the human workers' uncertainty and unpredictability, a framework needs to be proposed to achieve a safe and efficient human-swarm collaboration.

\subsection{Map Setting and Symbol definition} \label{map}

As shown in Figure 1, the physical footprint of each agent is a circle with a radius of ${R}$. At each minimum time step, the agent is located in the center of a square grid with a side length of $2{R}$. The minimum time step $\phi$ is defined as the time it takes for an agent to move at a constant velocity $v$ from the center of one square grid of side length $2{R}$ to the center of another adjacent grid. {\color{blue}Within a minimum time step, an agent can only move to one of the four adjacent grids from its original grid or stay in the original grid. Also, each workstation occupies only one grid in the map.} The PD tasks are completed when the agent arrives and occupies the corresponding grid for a prescribed time interval.
{\color{blue}In the (HS)$_2$, human workers and robots are placed in the same space without setting their own work space, and they complete the same batch of orders. Robots can go where humans can go. However, due to safety requirements, when the robot detects that there are human beings in a safe distance on the road ahead, the robot will stop in an emergency. Similarly, humans and robots have a cooperative relationship and will not deliberately approach the safe distance of the robot to block its movement.}

\begin{table}[h]
\caption{Notation and variables used in the mathematical model.}
\label{tab2}
\begin{tabular}{ll}
\hline
$\mathcal{R}$                       & Robot number set                     \\
$\mathcal{H}$                       & Human number set                      \\
$\mathcal{P}$                       & Agent (human and robot) initial position set          \\
$\mathcal{T}_p$     &Pickup task number set\\
$\mathcal{T}_d$     &Delivery task number set\\
$\mathcal{S}_p $                      & Pickup task starting point set             \\
$p_w $                    & Warehouse of pickup tasks            \\
$\mathcal{S}_d $                      & Delivery task origin set              \\
$\mathcal{G}_d $                      & Delivery task destination set               \\

$\mathcal{M}$                       & {\color{blue}Grids set in the global map as the work scenario}              \\
$\mathcal{M}_w$   &{\color{blue}The set composed of all white grids forming possible roads}\\
$\mathcal{M}_b$   &{\color{blue}The set composed of all black grids that can issue tasks as workstations}\\
$\mathcal{M}_g$   &{\color{blue}The set of green grids that composed of charging places}\\

$\mathcal{V}$                       & {\color{blue}Set of task nodes, agent nodes,}  {\color{blue}and a warehouse node } \\
$\mathcal{E}$                       & Set of edge between two nodes in  node set $\mathcal{V}$      \\
$\mathcal{W}$                       & {\color{blue}Set of passable grids in map $\mathcal{M}$, task nodes, } {\color{blue}and agent nodes} \\
$\mathcal{A}$                       & Set of edge between two nodes in node set $\mathcal{W}$     \\
$\delta_{a,M}$       & Binary decision variable that represents whether agent $a$ is in mode $M$              \\
$E_a$        &  The current remaining power of agent $a$ \\
$E_{set,a}$  &  The minimum working power of agent $a$  \\
$M_s$        &A very large positive constant. \\
$x_{i,j}^a$     & Binary decision variable that represents whether agent $a$ ($r$ or $h$) in pickup mode go from node $i$ to $j$   \\
$y_{i,j}^a$   &  Binary decision variable that represents whether agent $a$ ($r$ or $h$) in delivery mode go from node $i$ to $j$   \\
$t^h_j$    & Time to reach the start of the task  \\
$t^w_j$    & The waiting time of robot $i$ at node $j$  \\
$t^b_j$    & The opening time in the time window of task $j$                   \\
$t^e_j$    & The closing time in the time window of task $j$                  \\
$t^s_j$    & The length of time required to load or unload for task $j$  \\
$\omega^a_{i,j}$ &The cost for executing task from node $i$ to node $j$ of the agent $a$ ($r$ or $h$)  \\
$d_{i,j}$  & the journey distance from node $i$ to node $j$ \\
$m_j^a$    & Binary value that represents whether task $j$ must be completed by agent $a$ ($r$ or $h$) \\
$p_a(t)$      &The position coordinates of agent $a$ at time $t$  \\
$T$     & The update time step  \\
$T_D$     & The model prediction horizon  \\
$l_j$   & The weight of the cargo at task $j$ \\
$L_a$   & The maximum cargo capacity of agent $a$ ($r$ or $h$)  \\
$\Pi_a$ &The ordered set of tasks assigned to agent $a$ ($r$ or $h$)\\
$z_{i,j}^r $   &  {\color{blue}Binary decision variable that represents whether}  {\color{blue}robot $r$ move from node $i$ to $j$ in node set $\mathcal{W}$}\\
$P_r^{\tau}$  &The set of path planning for robot $r$ to complete task $\tau$ \\
$\pi$  &The set of PF results for the robot swarm \\
\hline
\end{tabular}
\end{table}

{\color{blue}}

There are a total of $n+m$ tasks as $\mathcal{T}$, consisting of $n$ pickup mode tasks as $\mathcal{T}_p$ and $m$ delivery mode tasks as $\mathcal{T}_d$ , and $k+b$ agents, consisting of $k$ robots and $b$ human workers. {\color{blue}Robots and human workers work together to transport goods in a scenario $\mathcal{M}$. The scenario $\mathcal{M} = \mathcal{M}_b\cup \{p_w\}\cup \mathcal{M}_g\cup\mathcal{M}_{w}$, indicating that the scenario consists of workstations where tasks can be published, warehouses where goods can be stored, robot charging stations, and roads for agents to travel.}
The positions of a robot and a human at time $t$ are defined as $p_r(t)$ and $p_h(t)$, abbreviated as $p_r$ and $p_h$, both are tuples.
The starting point set of the pickup mode tasks $\mathcal{T}_p$ is defined as $\mathcal{S}_p=\{s_i\}_{i \in \textbf{I}[1,n]}$, which are tuples. The goods at the starting points need to be transported to the same warehouse $p_w$. The origin set of the delivery mode tasks is defined as $\mathcal{S}_d=\{s_j\}_{j \in \textbf{I}[n+1,n+m]}$, the delivery mode tasks $\mathcal{T}_d$ must have definite destination set $\mathcal{G}_d=\{g_j\}_{j \in \textbf{I}[n+1,n+m]}$, which are tuples.

The first layer abstracts the grid graph as a weighted directed point-line graph $\mathcal{G}_1=(\mathcal{V}, \mathcal{E})$, where $\mathcal{V}=\{p_a, s_i, s_j, g_j, p_w \}_{a \in \textbf{I}[1,k] \cup \textbf{I}[k+1,k+b], i \in \textbf{I}[1,n], j \in \textbf{I}[n+1, n+m]}$, and $\mathcal{E}$ is a set composed of directed edges $(i,j)$, with $i, j \in \mathcal{V}$. At each time $t$, each agent either serves as a vertex, waits at a vertex, or moves to a vertex. In any of these cases, the request associated with the vertex can be known at time $t$. {\color{blue}We} will complete the TA in pickup mode and delivery mode on a larger time scale $T_D$. The second layer uses a more detailed grid map $\mathcal{G}_2$ to abstract as an unweighted directed point-line graph $\mathcal{G}_2=(\mathcal{W}, \mathcal{A})$, where all passable grids  ${w_i}$ with $i \in \textbf{I}[1,N]$ ($N$ represents the total number of passable nodes in the grid map $\mathcal{M}$), in the map form a node set $\mathcal{W}$, and $\mathcal{A}$ is the set formed by all the edges $(i,j),~\forall i,j \in \mathcal{W}$. We will complete the PF of pickup mode tasks and delivery mode tasks in a smaller time scale.

Meanwhile, the meaning of almost all the symbols in this article are listed in {\color{blue}Table \ref{tab2}}.

\subsection{Basic settings} \label{setting}
In the detailed \textbf{Problem (HS)$_2$PD} processing, we mainly consider the following aspects:
\begin{enumerate}
    \item Herein we consider a (HS)$_2$ with $k$ robots and $b$ human workers. Each task must be assigned to only one agent (robot or human).
    \item Each task has an opening time and a closing time, which are the earliest time to be picked up and the latest time to be picked up, respectively. The period from the opening time to the closing time of a task is called \textbf{time window}. The goods of each workstation can only be loaded within the time window specified by its task.
    \item The loads occupied by the tasks of workstations can be different but are fixed with an upper bound. \textbf{The limited loading capacities} of agents can also be different but have a lower bound. {\color{blue}This means that the limited loading capacity of a human can be different from that of a robot.}
    \item We assume that the agent can perform forward, backward, left, right, and stop actions without considering specific kinematic constraints. Assume that the movement costs in the four directions of front, back, left, and right are the same both for human workers and robots.
    \item  The emergence and disappearance of task requests are \textbf{dynamic} and \textbf{unpredictable}, which may be caused by the release and cancellation of tasks by customers. Moreover, the unreliability of human force may also lead to this unpredictable task emergence.

\end{enumerate}

Since in the work environment, a (HS)$_2$ is always used to complete multiple tasks with different requirements, such as the TA of the worker bee colony in the hive. Their functions can be changed according to the changes in environmental conditions and the needs of bee colony. According to the different task requirements and the charging requirements of the robot, this article reasonably defines three operating modes: pickup task mode, delivery task mode, and charging mode, respectively.
\begin{enumerate}
    \item[1)] In the \textbf{pickup task mode}, an agent will be assigned to an ordered task set, and then the intention is to arrive at the corresponding task starting points for pickup in turn. By completing pickup, the goods picked up at the workstations are transported to a fixed warehouse $p_w$ for subsequent processing. We only care about how the agent moves the goods from the workstation to the warehouse, not the specific processing in the warehouse.
    \item[2)] In the \textbf{delivery task mode}, an agent is assigned a delivery task, then goes to the origin of the task for loading goods, and transports the goods to another paired destination to unload in the first time directly. An agent can only respond to a delivery task request from one workstation at a time.
    \item[3)] In \textbf{charging mode}, the robots return to the nearest charging point before the battery runs out, and human workers go to the rest area outside the map to rest for a while.
\end{enumerate}

\begin{figure}[thpb]  
	\centering 
\includegraphics[scale=0.4]{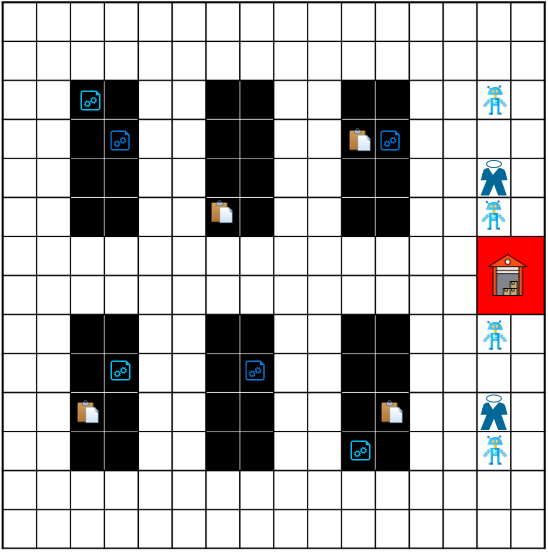} 
 	\caption{A map with $16 \times 14$ grids: Green and blue lines represent possible routes for agents to complete pickup mode tasks and delivery mode tasks, respectively. {\color{blue}The red line represents a possible route for a robot going to charge.}}
	\label{fig1} 
\end{figure}

   The possible trajectories of these three modes are shown in Figure \ref{fig1}.

\begin{remark}\label{remark1}
    {\color{blue}Delivery mode tasks have paired starting and destination locations. An agent that accept a delivery mode task cannot access other task requests after it arrives at the starting location. The agent responds to other pickup mode tasks only after completing the currently assigned delivery mode task. However, the starting point and destination of the pickup mode task are not paired. So when the load of a agent that accept pickup mode tasks has not reached its limited load capacity, it can still respond to the delivery mode task request of the workstation.}

\end{remark}

\vspace{0.2cm}

\section{Main Results} \label{sec3}

{\color{blue}In this section, we propose a solution framework for the requirements and characteristics of the \textbf{Problem (HS)$_2$PD} described in the previous section.}

It is pointed out in \cite{khamis2015multi} that how to best allocate various tasks with different requirements and constraints to heterogeneous unreliable robots with different capabilities is a complicated problem.
Meanwhile, in \textbf{Problem (HS)$_2$PD}, human workers are generally at a loss to respond to a large number of task requests to be assigned. However, under a known environment, ordinary human workers can easily find a feasible path {\color{blue}on} the map. Therefore, \textbf{Problem (HS)$_2$PD} can be divided into two sub-problems, that is, TA (TA) problem and PF (PF) problem. In the TA problem, the characteristics of human fatigue, low loading capacity, autonomy, and uncertainty are fully considered in order to form a better human-swarm TA scheme and make the (HS)$_2$ more efficient. In the PF problem, {\color{blue}the feasible path scheme of the robot swarm is formed, and the obstacle avoidance between robots and human workers is considered so that the (HS)$_2$ can meet the security.} Inspired by \cite{fu2021real}, we propose a two-layer control structure to solve the TA and PF problems of the (HS)$_2$ in turn. The first and second layers can be described as two different mixed-integer programming (MIP) problems and solved by various standard {\color{blue}softwares}. At the same time, using a two-layer optimization scheme to solve \textbf{Problem (HS)$_2$PD} can reduce the computational complexity.

\subsection{Two-layer architecture and dynamic allocation strategy}
Our strategy is to classify tasks with different requirements visually. In (HS)$_2$, {\color{blue}each agent can complete a task in a specified mode with a corresponding cost of movement.} This requires us to consider and determine the following factors: a) what type of task an agent is going to perform, b) to which one of these types of tasks the agent is assigned, and c) finally, how safely and efficiently the agent moves and performs the tasks it receives. Therefore, for factors a) and b), we design a two-layer optimization scheme to solve the TA problem in the first layer, which is divided into pattern allocation and task selection, and for factor c), the second layer of the two-layer optimization scheme is responsible for solving the exact PF problem. See Table \ref{tab1} to understand the detailed framework for solving the \textbf{Problem (HS)$_2$PD}.
Specifically, in the top-level problem, if the agent's mode assignment result is determined, then the bottom-level delivery mode TA and pickup mode task planning can be separated and solved as two independent planning problems.


\begin{table*}[]
\caption{{\color{blue}Two-layer} optimization scheme for \textbf{Problem (HS)$_2$PD}}
\label{tab1}
\resizebox{\textwidth}{!}{
\begin{tabular}{@{}ccccc@{}}
\toprule
Layers                       & Constituent elements               & Formulated problem                                    & Obtained results \\ \midrule
\multirow{3}{*}{First layer} & \multirow{3}{*}{Humans and robots} & Top level of bi-level programming                     & Mode allocation \\ \cmidrule(l){3-4}
                             &            & \multirow{2}{*}{Bottom level of bi-level programming} & Pickup mode assignment   \\ \cmidrule(l){4-4}
                             &                          &                                                       & Delivery mode assignment  \\ \midrule
Second layer                 & Robots                             & Mixed-integer programming                       & Detailed PF      \\ \bottomrule
\end{tabular}}
\end{table*}

\begin{figure}[thpb]
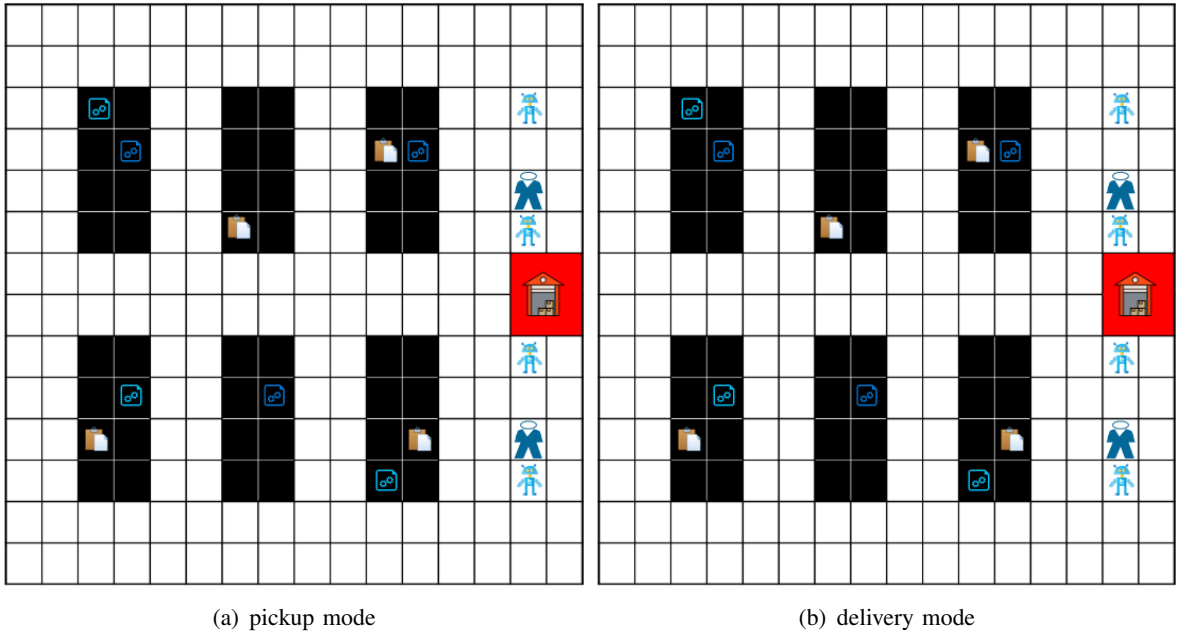
  
	\centering 
\subfigure[pickup mode]{\label{fig1a}\includegraphics[scale=0.4]{change.png}}
\subfigure[delivery mode]{\label{fig1b}\includegraphics[scale=0.4]{change.png}}
 	\caption{{\color{blue}Pickup tasks and the delivery tasks to be executed by human workers and robots.}}
	\label{fig2} 
\end{figure} 
In the charging mode, the robot is called back to the nearest charging point, and the human will go to rest. The agent in charging mode navigates autonomously, without allocating tasks and calculating paths. Next, we will introduce the pickup and delivery tasks in detail.

In the pickup mode, as shown in Figure \ref{fig1a}, one agent is assigned to an ordered task set $\Pi$, which means that the agent needs to patrol the task starting point pickup of these tasks in turn, and finally patrol to the fixed warehouse. So if an agent is in the pickup task mode, its movement can be simulated as an agent arriving in the center of several grids in a grid map. The first layer in the two-layer optimization scheme determines the ordered set of tasks assigned to each agent in the pickup task mode, only determines its rough patrol route, and does not care about the specific movement trajectory of the agent for the time being. Only the patrol trajectories of robots are determined by a PF algorithm in the second layer. After obtaining the TA sequence, the human workers will find the path autonomously. By implementing TA and rough route and detailed route planning in a two-layer optimization scheme, decoupling is realized to pursue less computational complexity.

In the delivery mode, as shown in Figure \ref{fig1b}, the agent will only get a delivery mode mission, go to the origin of the mission for pickup, and then immediately go to the destination of the task for delivery, which means we only need to think about assigning the origins of the delivery mode tasks in TA problem. The first layer of the two-layer optimization scheme determines the delivery task assigned by the agent of each delivery task mode. Similar to the pickup task mode, the specific path of the robot in the delivery task mode is also through the second layer of the two-layer optimization scheme to obtain.

In a dynamic environment, the number of tasks and processing {\color{blue}states} change in real-time, which is unknown and cannot be a priori. In general, in practice, the task list may have changed several times by the time an agent executes a delivery mode task, given that the agent may execute {\color{blue}the} delivery mode task for longer than the pickup mode task. The task list may change because an old task is {\color{blue}canceled}, a new task is released, or a human unknowingly performs a task, that is not his, resulting in a waste of resources for the robot that was expected to handle the task. To solve these dynamic problems and in order to avoid short-term delivery decisions, the receding horizontal strategy \cite{schouwenaars2004receding} prediction decision method with moving prediction window is adopted. We set a prediction horizon $T_D$ for TA of (HS)$_2$, whose size is several update time step $T$. The statement in remark \ref{remark1} that the agent executing the delivery mode task cannot change the task mode midway is reasonable since the update time step is chosen based on the time it takes to execute a delivery mode task.
The first layer programming problem will solve the mode assignment and TA problem within a prediction horizon $T_D$, and get an optimal decision sequence. However, the (HS)$_2$ only executes the decision within the first update time step $T_1$, that is,  the first item of the decision sequence. The task changes that occur within this update time step will not directly affect the assigned task set, but will be kept and recorded and will only be considered in the next update time step $T_2$ to optimize the decision sequence again. Even if the actual task situation changes frequently in a short period of time, it will not be responded to immediately. These changes will be processed after the current update time step has passed. In other words, the extent of each allocation calculation is a prediction horizon. During an update time step, the first-layer programming mode and TA results do not change. However, according to the current environment, the allocation decision is updated once per the update time step.


\begin{figure}[thpb]  
	\centering 
	\includegraphics[scale=0.6]{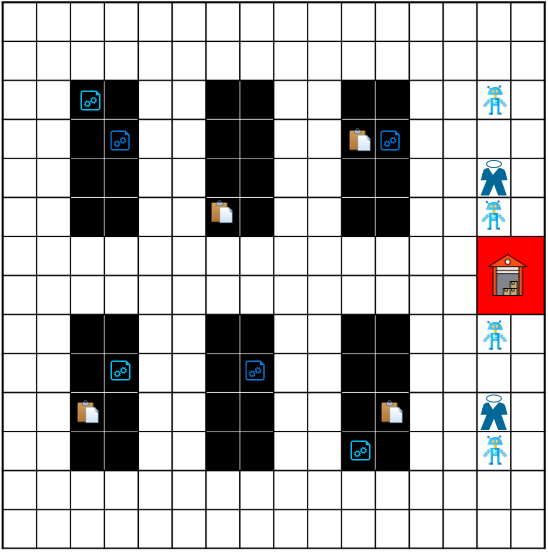} 
 	\caption{Visualization of the time span of an agent for pickup mode tasks.}
	\label{fig4} 
\end{figure} 

\subsection{First level: Task Allocation}\label{TA}
In an environment where robots and human workers work together, robots and human workers are equivalent to two kinds of heterogeneous agents that participate in TAs together. However, considering the subjectivity of human workers in the actual working environment, if they are assigned to those that they need to walk a long distance. These tasks will significantly accelerate human workers’ fatigue, which will affect the quality of human workers’ work. On the other hand, some tasks have special attribute requirements and can only be completed by robots or human workers.

In this section, we turn the assignment problem of the agent running pattern into a mixed-integer bi-level programming problem. The optimization goal is to minimize the cost of completing all tasks, and the formula of mode assignment is as follows:
\begin{align}
    &\min_{\delta} \ \{\min_{x,y} \ \{F(\delta,x,y)\}\} ,\label{MA}\\
    {\rm{s.t.}} \quad &\sum_{\mathrm{M} \in \mathcal{M}_m} \delta_{a,\mathrm M} = 1,~\forall a \in \mathcal{R}\cup\mathcal{H}, \label{m2}\tag{\ref{MA}{a}}\\
    &\delta_{a,3} E_{a}\leq E_{set,a},~\forall a \in \mathcal{R}\cup\mathcal{H}\label{m3}\tag{\ref{MA}{b}}\\ ~ &(1-\delta_{a,3})E_{set,a} \leq E_a,~\forall a \in \mathcal{R}\cup\mathcal{H},\label{m4}\tag{\ref{MA}{c}}\\
    &x \in \mathrm{\Phi}_1(\delta) ,~y \in \mathrm{\Phi}_2(\delta),\label{m5}\tag{\ref{MA}{d}}
\end{align}
where $F(\delta,x,y)= C_{P}(\delta,x) + \gamma C_{D}(\delta,y)$, and $C_{D}(\cdot)$ and $C_{P}(\cdot)$ respectively represent the waiting cost of delivery tasks and the waiting cost of pickup tasks (see below for details) , parameter $\gamma \geq 0$ is the weight coefficient, which represents the relative importance of pickup tasks relative to delivery tasks. The mode assignment decision variable $\delta_{a,\mathrm M}$ is a binary variable indicating if the agent $a$ ($r$ or $h$) is assigned to mode $\mathrm M$, where $\mathrm M \in \mathcal{M}_m =\textbf{I} [1,3]$, corresponding to pickup task mode, delivery task mode and charging model , respectively. Distinguish between human workers and robots, $\delta:=[\delta_{r,\mathrm M}\ \delta_{h,\mathrm M}]^\mathrm{T}$. 
The TA decision variables $x$ and $y$ are the vectors of pickup task variables and delivery task variables defined in subsections \ref{norm_TA} and \ref{real_TA}, respectively.

Constraint (\ref{m2}) means that an agent can only be assigned to one operating mode. Constraint (\ref{m3}) and (\ref{m4}) mean that when the power $E_r$ of the robot is less than the predetermined threshold $E_{set,r}$ or the endurance $E_h$ of the human is less than the least energy $E_{set,h}$, the robot is assigned to the charging mode or a human will rest. In formula (\ref{m5}), $\mathrm{\Phi}_1(\delta)$ and $\mathrm{\Phi}_2(\delta)$ define the feasible set of the variable vector $y_1$ in the pickup mode tasks and the variable vector $y_2$ in the delivery mode tasks, respectively. When the result of the high-level mode assignment is determined, the low-level TA can be regarded as two independent MIP problems.

\subsubsection{Pickup Mode Allocation}\label{norm_TA}
Recall the point-line graph $\mathcal{M}_p=(\mathcal{V}, \mathcal{E})$ defined in subsection \ref{map}, we define $\mathcal{S}_p=\{s_i\}_{i \in \textbf{I}[1,n] }$ to represent the set of task starting points for all pickup mode tasks, and $\mathcal{P}=\{p_r,p_h\}_{ r \in \textbf{I}[1,k] , h \in \textbf{I}[1,b]}$ to represent the set of current positions of all agents.
The service time window of task $j$ is $(t^b_j, t^e_j)$. 
A visualization of the time span of a set of tasks is shown in Figure \ref{fig4}.

Then we can describe this MIP problem. The objective function of the pickup mode task as the total cost consisting of the total cost of completing tasks with robots and human workers and the extra cost for tasks that have released requests but no response:
\begin{figure*}[!]
    \begin{align}
	& \min_{x} \  C_{P}(\delta,x)  \label{TAP}\\
   {\rm{s.t.}}\ \ \  &  x_{i,j}^r = \{0,1\} ,  ~x_{i,j}^h = \{0,1\}, ~\forall r \in \mathcal{R},~ h \in \mathcal{H},~ (i,j) \in \mathcal{E},\label{decision variables} \tag{\ref{TAP}{a}}\\
  &\sum _{(i,j)\in \mathcal{E}} x_{i,j}^a \delta_{a,1} \leq 1 , ~\forall a\in \mathcal{R}\cup\mathcal{H},~i \in \mathcal{P},\label{sub_rob_out}\tag{\ref{TAP}{b}}\\
 &\sum _{a \in \mathcal{R}\cup\mathcal{H}}\sum _{(i,j)\in \mathcal{E}} x_{i,j}^a \delta_{a,1} \leq 1 ,~ \forall j \in \mathcal{S}_p,\label{sub_task_in}\tag{\ref{TAP}{c}}\\
 &\sum _{(i,j)\in \mathcal{E}} x_{i,j}^a \delta_{a,1} \leq \sum _{(j,i)\in \mathcal{E}} x_{j,i}^a \delta_{a,1} ,  ~ \forall a\in \mathcal{R}\cup\mathcal{H},~ i \in \mathcal{S}_p ,\label{sub_blance}\tag{\ref{TAP}{d}}\\
 &\sum _{(i,j)\in \mathcal{E}} \delta_{a,1}x_{i,j}^a = 0 ,~\forall a \in \mathcal{R}\cup\mathcal{H},~j \in \mathcal{P},\label{rob_blance}\tag{\ref{TAP}{e}} \\
 &M_{s}(\delta_{a,1}x_{i,j}^a -1) \leq t^h_i + t^w_i +{t^s_i} + t_{i,j}^a -t^h_j  \leq M_{s}(1- \delta_{a,1}x_{i,j}^a) , ~\forall a \in \mathcal{R}\cup\mathcal{H},~ (i,j) \in \mathcal{E} ,\label{time_limit}\tag{\ref{TAP}{f}}\\
 &0 \leq t^b_j \leq  t^h_j \leq t^e_j,~ \forall j \in \mathcal{S}_p,\label{time window}\tag{\ref{TAP}{g}}\\
 & 0  \leq \delta_{a,1}t^h_j \leq T_D, ~\forall j \in \mathcal{S}_p,\label{forecast horizon}\tag{\ref{TAP}{h}}\\
  &\sum _{(i,j)\in \mathcal{E}} \delta_{a,1}x_{i,j}^a l_j \leq L_a  ,~\forall a \in \mathcal{R}\cup\mathcal{H}, ~i \in \mathcal{S}_p\cup \mathcal{P},~j \in \mathcal{S}_p,\label{load}\tag{\ref{TAP}{i}},
    \end{align}
    where
\begin{align}
 C_{P}(\delta,x) = \sum_{r \in \mathcal{R} }\sum_{(i,j)\in \mathcal{E}} \omega_{i,j}^r \delta_{r,1} x^r_{i,j} &+  \sum_{h \in \mathcal{H} }\sum_{(i,j)\in \mathcal{E}} \omega_{i,j}^h \delta_{h,1} x^h_{i,j} \nonumber\\  & + \sum_{{i\in \mathcal{P},j \in \mathcal{S}_p}} (\max\{\omega_{i,j}^r,\omega_{i,j}^h\} + \max\{\omega_{j,p_w}^r,\omega_{j,p_w}^h\})\cdot\delta_{a,1} (1-\sum_{a \in \mathcal{R}\cup \mathcal{H} }x^a_{i,j}),\label{cp_detail} 
\end{align}
    \hrulefill
\end{figure*}

and $x^a_{i,j}:=[x^r_{i,j}\ \  x^h_{i,j}]^\mathrm{T}$. The terms $\omega^r_{i,j} , \omega^h_{i,j} > 0 $ represent the costs for executing task from node $i$ to node $j$ on behalf of the robot and of the human, respectively, and
\begin{align}
   &\omega_{i,j}^r = \alpha d_{i,j} + M_s (1-m_j^r),& \label{weight_r}\\
   &\omega_{i,j}^h = (1-\alpha)d_{i,j} + M_s (1-m_j^h),&\label{weight_h}
\end{align}
where $0<\alpha<1$ represents the weight of the journey cost of the robot relative to the human from node $i$ to node $j$, and $m_j^r$ and $m_j^h$ are binary values representing whether task $j$ must be completed by robots or human workers.

{\color{blue}In the design of objective function, we consider the difference of working characteristics between human and robot. Robots don't get tired while working, but human workers tend to be lazy and tend to respond to tasks that require less effort (close to them). In Equations (\ref{weight_r}) and (\ref{weight_h}), when the weight $\alpha < 0.5$, it will cost more for human to accept a long-distance task. Therefore, our proposed framework would tend to assign humans closer tasks to reduce human fatigue and increase human productivity. On the other hand, the limited loading capacity of human is smaller than that of robot, so assigning more tasks to robot can increase the completion efficiency. In Equation (\ref{cp_detail}), the last item is the additional cost of the task not being assigned to any agent. Therefore, our proposed framework would tend to assign more tasks to robots to make the total cost of (HS)$_2$ smaller.}


Constraint (\ref{decision variables}) means that the decision variable is a binary value.
Constraint (\ref{sub_rob_out}) requires that each agent in {\color{blue}the} pickup model must start from his current location or wait in place.
Constraint (\ref{sub_task_in}) requires at most one agent to be assigned to each pickup task.
Constraint (\ref{sub_blance}) requires the flow balance of the task starting point of each pickup task.
Constraint (\ref{rob_blance}) requires flow balance at the current position of the robot.
Constraint (\ref{time_limit}) requires a certain travel time $t^h_i$ for the agent to go to the starting point of each task and a certain service time $t^s_i$ for pickup at the starting point of the task. When the agent arrives before the opening time of the task, he will wait here for the task to be released. {\color{blue}Note that the travel time $t^a_{i,j}$ here should take into account the walking speeds of human workers and robots, road congestion, and obstacle bypass time.}
Constraint (\ref{time window}) requires the agent to arrive at the starting point of the task within the time window $(t^b_i,t^e_i)$.
Constraint (\ref{forecast horizon}) requires that the time for the agent to reach the starting point of the task cannot exceed the predicted horizon $T_D$.
Constraint (\ref{load}) requires that the total amount of goods picked up by the agent $a$ at the starting point of each task is less than the maximum load $L_a$.

Then the feasible set can be defined by the constraints, and the optimization vector is :
\begin{align*}
    x=[x^r_{i,j}\ \  x^h_{i,j}\ \  t^h_j]^ \mathrm{T}_{r \in \mathcal{R}, h \in \mathcal{H}, i \in \mathcal{S}_p\cup \mathcal{P}, j \in \mathcal{S}_p}.
\end{align*}

\subsubsection{Delivery Mode Allocation}\label{real_TA}
Similar to the pickup task mode, describe the MIP problem of the delivery task mode:

\begin{figure*}
    \begin{align}
    &\min_{y} \  C_{D}(\delta,y) \label{TAD}\\
 {\rm{s.t.}} \quad\ &\sum _{(i,j)\in \mathcal{E}} y_{i,j}^a\delta_{a,2} \leq 1,~\forall a\in \mathcal{R}\cup\mathcal{H},~i \in \mathcal{P},\label{rea_rob_out}\tag{\ref{TAD}{a}}\\
&  \sum _{a \in \mathcal{R}\cup\mathcal{H}}\sum _{(i,j)\in \mathcal{E}} y_{i,j}^a\delta_{a,2} \leq 1 ,~ \forall j \in \mathcal{S}_d,\label{rea_task_in}\tag{\ref{TAD}{b}}\\
  &0 \leq t^b_j\leq t^h_j \leq t^e_j,~\forall j \in \mathcal{S}_d,\label{rea_time_window}\tag{\ref{TAD}{c}}\\
   &\sum _{(i,j)\in \mathcal{E}} \delta_{a,2}y_{i,j}^a = 0 ,~\forall a \in \mathcal{R}\cup\mathcal{H},~j \in \mathcal{P},\label{rea_rob_blance}\tag{\ref{TAD}{d}}\\
  & 0 \leq \delta_{a,2}t^h_i \leq T_D, ~\forall i \in \mathcal{S}_d,\label{R_forecast horizon}\tag{\ref{TAD}{e}}
\end{align}
where
\begin{align}
   C_{D}(\delta,y) = \sum_{r \in \mathcal{R} }\sum_{(i,j)\in \mathcal{E}} \omega_{i,j}^r \delta_{r,2} y^r_{i,j} &+  \sum_{h \in \mathcal{H} }\sum_{(i,j)\in \mathcal{E}} \omega_{i,j}^h \delta_{h,2} y^h_{i,j} \nonumber\\&+ \sum_{i\in \mathcal{P},j \in \mathcal{S}_d} \max\{\omega_{i,j}^r,\omega_{i,j}^h\} \delta_{a,2} (1-\sum_{a \in \mathcal{R}\cup \mathcal{H} }y^a_{i,j}),\nonumber
\end{align}
    \hrulefill
\end{figure*}

and $y^a_{i,j}:=[y^r_{i,j}\ \  y^h_{i,j}]^\mathrm{T}$. The set $\mathcal{S}_d=\{s_i\}_{ i \in \textbf{I}[n+1,n+m] } $ represents the set of origins of all delivery mode tasks, and $\mathcal{G}_d=\{g_j\}_{ j \in \textbf{I}[n+1,n+m]}$ represents the set of destinations of all delivery mode tasks.


Constraint (\ref{rea_rob_out}) requires that each agent in {\color{blue}the} delivery tasks model must start from his current location or wait in place.
Constraint (\ref{rea_task_in}) requires at most one agent to be assigned to each delivery task.
Constraint (\ref{rea_time_window}) requires the agent to arrive at the origin of the task within the time window $(t^b_i,t^e_i)$.
Constraint (\ref{rea_rob_blance}) requires flow balance at the current position of the agent.
Constraint (\ref{R_forecast horizon}) requires that the time for the agent to reach the origin of the task cannot exceed the predicted horizon $T_D$.

Then the feasible set can be defined by the constraints, and the optimization vector is:
\begin{align*}
    y = [y^r_{i,j}\ \  y^h_{i,j}\ \ t^h_j]^ \mathrm{T}_{r \in \mathcal{R}, h \in \mathcal{H}, i \in \mathcal{P},j \in \mathcal{S}_d}.
\end{align*}
\subsection{Second Level: Path Finding} \label{PF}
Through mode assignment, the agent's operating mode $\delta_{a,M}$ in a prediction horizon and the ordered set of tasks $\Pi_a = \{\tau_1, \tau_2,...\}$ assigned to it are determined. Next, we will solve the PF problem of the second layer for robots on this basis.

In order to facilitate PF, we believe that after the global map is rasterized, several vertically intersecting two-way passable roads will be formed, which ensures that each service point has a passable road to reach \cite{fransen2020dynamic}. {\color{blue}Only the TA plan within the forecast range is considered in the TA section, and the TA plan is updated at each update time step. Our PF strategy is to decouple the ordered task set of each robot into multiple single-destination PF problems in turn. This processing method can also adapt to the task requirements of two different modes. We assume that each robot can obtain the task assignment plan and path plan of all robots from the cloud.}

Recall the point-line graph $\mathcal{G}_2=(\mathcal{W}, \mathcal{A})$ defined in subsection \ref{map}. For the robot $r$ to perform the task $\tau, \tau \in \Pi_r$, we define the position of the workstation where the task $\tau$ is located as $g$, so correct the point set $\mathcal{W}=\{w_i\}_{i \in N}\cup\{p_r\}\cup\{g\}$. Then the MIP problem of the second layer can be described as:
\begin{figure*}
    \begin{align}
  &\min_{z} \  C_Z(z)  \label{PFA}\\
 {\rm{s.t.}} \quad &z_{i,j}^r = \{0,1\} ,~\forall r \in \mathcal{R},~(i,j) \in \mathcal{A},\label{z decision variables}\tag{\ref{PFA}{a}}\\
  &\sum _{(p_r,j)\in \mathcal{A}} z_{p_r,j}^r = 1 ,~\forall r\in \mathcal{R},\label{rob_out}\tag{\ref{PFA}{b}}\\
  & \sum _{(i,g)\in \mathcal{A}} z_{i,g_r}^r = 1 ,~ \forall r \in \mathcal{R},\label{ending_in}\tag{\ref{PFA}{c}}\\
  &  \sum _{(i,j)\in \mathcal{A}} z_{i,j}^r = \sum _{(j,i)\in \mathcal{A}} z_{j,i}^r ,~ \forall r\in \mathcal{R}, ~j \in \mathcal{W} \backslash \{p_r\} ,\label{blance}\tag{\ref{PFA}{d}}\\
 &  p_r(t) \neq p_b(t) ,~\forall r,b \in \mathcal{R} ,~\forall t\geq0 ,\label{collision1}\tag{\ref{PFA}{e}}\\
 & (p_r(t),p_r(t+\phi)) \neq (p_b(t+\phi),p_b(t)) ,~\forall r,b \in \mathcal{R},~\forall t\geq0 ,\label{collision2}\tag{\ref{PFA}{f}}\\
& M_{s}(z_{i,j}^r -1) \leq t^r_i +{t^w_i} + t_{i,j}^r -t^r_j  \leq M_{s}(1- z_{i,j}^r) ,~\forall r \in \mathcal{R}, (i,j) \in \mathcal{A} ,\label{path_time_limit}\tag{\ref{PFA}{g}}
\end{align}
where
\begin{align}
 C_Z(z) =  \sum_{(i,j)\in \mathcal{A}}d_{i,j} z^r_{i,j} 
 .\nonumber
\end{align}
    \hrulefill
\end{figure*}

Constraint (\ref{z decision variables}) means that the decision variable if the robot $r$ go from node $i$ to $j$ is a binary value.
Constraint (\ref{rob_out}) requires that each robot must start from {\color{blue}its} current location or wait in place.
Constraint (\ref{ending_in}) requires that there must be one robot at the ending point of task $j$.
Constraint (\ref{blance}) requires the flow balance of the path set of each task.
Constraints (\ref{collision1}) and (\ref{collision2}) require that there is no collision between any two robots.
Constraint (\ref{path_time_limit}) is the time stream limit.

The order in which each robot independently executes its respective individual task with MIP follows the order of closing time of the time window. The solution result of the mixed-integer programming problem is an ordered set of passing nodes, which constitute the path set $P_r^{\tau}=\{\widetilde{p_r}(t)\}$ for the robot $r$ to complete the task $\tau$, where $\widetilde{p_r}(t)$ represents the node position that robot $r$ will occupy at time $t$ in the planning result. This result $P_r^{\tau}$ is passed to subsequent MIP problems as the $p_b(t)$ term in constraints (\ref{collision1}) and (\ref{collision2}). This operation ensures that there will be no conflicting results between multiple distributed MIP problems.



Next, the result is handed off to the underlying robotic hardware for execution with the minimum time step $\phi$. Assuming that the motion step is $2R$ and the robot moves at a constant speed $v_r$, then the minimum time step $\phi=2R/v_r$. The motion model of the robot can be expressed as follows:
\begin{align}
    p_r(t+\phi) = p_r(t)+ {2R} \begin{bmatrix}\cos({\theta(t)})&\sin({\theta{(t)}})\end{bmatrix}^{\rm T}&,\nonumber
\end{align}
where $p_r(t)$ is the position coordinate of robot $r$ at time $t$. After a minimum time step $\phi$, it will reach the next position coordinate $p_r(t+\phi)$. $\theta(t) \in \Theta=\{{0},{\frac{\pi}{2}},{\pi},{-\frac{\pi}{2}}\}$ means the moving direction(forward, backward, or 90-degree turn) that the robot $r$ needs to perform in the next time step at time $t$. The specific motion decision and the entire solution process of the \textbf{Problem (HS)$_2$PD} can be seen in Algorithm \ref{HSHSPD} for detail.

{
\begin{algorithm}[t]
  \caption{An algorithm for \textbf{Problem (HS)$_2$PD}} 
  \label{HSHSPD}
  \raggedright {\bf Input:} 
  Agent (human and robot) set: $\mathcal{H} \cup \mathcal{R}$. The current remaining power of agent $a$: $E_a$.  Tasks set: $\mathcal{T}$ . Point set in the map: $\mathcal{M}$ . The update step size $T$.


  \begin{algorithmic}[1]

  \State $t=0$
  \For{tasks not fully completed}
    \If{$t = N \cdot T$}
        \State $\pi \gets \varnothing$
        \State $Update(\mathcal{T},\mathcal{R},\mathcal{H}$)
        \State $\Pi^* \gets Task-assignment (\mathcal{T},\mathcal{R},\mathcal{H})$ \Comment{An assignment that satisfies the constraints.}
        \State $\Pi_r \gets Select-robot-task-sequence (\Pi^*)$
        \For{each robot and each task}
            \State $P_r^\tau \gets Path-finding(\Pi_r,\mathcal{R})$
            \State $\pi=\pi \cup P_r^\tau$
            \Comment{Conflict-free paths.}
        \EndFor
        \State \hspace*{-0.05in} \textbf{endfor}
    \EndIf
    \State $Move (\pi$) \Comment{Follow the predetermined path for a minimum step size.}
    \State $t = t + \phi$
    %

  \EndFor
  \State \hspace*{-0.05in} \textbf{endfor}

  \Function {Move}{$\pi$}
    \If{Obstacles stop within two minimum time steps ahead}
        \State{The robot stop and wait};  \Comment{Guaranteed security.}
    \Else
        \State {Move forward a minimum time step as planned $\pi$}.
    \EndIf

  \EndFunction

  \end{algorithmic}
\end{algorithm}
}

\begin{remark}
    {\color{blue}Although the planned path $\pi$ is conflict-free between robots, it may conflict with human movements in the map. The conflict is inevitable since the whole environment is dynamic, there may be delays in communication between robots and the cloud, and human behavior is uncertain. But this conflict can be resolved within our framework. The security issues between robots and human workers is first satisfied by the sensor in the underlying hardware of robots. The robots will make an emergency stop when it detects a human workers in the path ahead and wait for the obstacle to be removed or eliminated.
    Then note that if the obstacle is always present, this will not cause the robot to go into a deadlock state, since at the next update time steps $T$, the TA and PF will be re-executed, resulting in a new conflict-free feasible path.}
\end{remark}

\section{Numerical Simulation}
\label{SecSm}

\begin{figure}[thpb]  
  \centering 
  \includegraphics[scale=0.6]{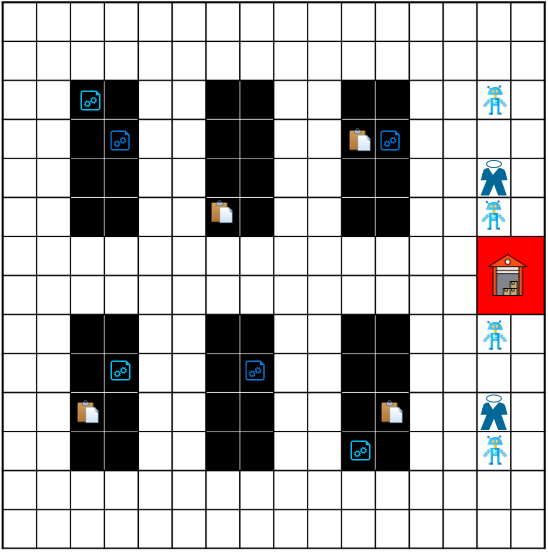} 
  \caption{A map with $16 \times 13$ grids: black tiles are static obstacles, white tiles are corridors, blue tiles represent potential origins and destinations (endpoints) of the tasks, and purple and yellow tiles represent starting locations of the robots and an agent.}
  \label{fig6} 
\end{figure} 

{\color{blue}In this section, we first conduct a simulation experiment on the \textbf{Problem (HS)$_2$PD} using our proposed method. The feasibility of our proposed method is verified. Secondly, several sets of simulation experiments are carried out under different problem scales. Some quantitative metrics are counted to assist in evaluating the effectiveness of the proposed method. Finally, through a comparison with xx algorithm, it is proved that our proposed controller has certain advantages.}
\subsection{A simulation experiment}
We perform our experiments on a 16 × 13 storage map as shown in Fig. \ref{fig6}, where black tiles are static obstacles, white tiles are corridors, blue tiles represent potential origins and destinations (endpoints) of the tasks, and purple and yellow tiles represent starting locations of the robots and {\color{blue}an} agent. The traversable roads in the map are all two-way traversable. In order to facilitate the realization of experimental programming and reduce the amount of calculation, the passable nodes are reduced to all intersections and tiles that must be passed to reach the destination workstation in the PF, which will not affect the PF results.


In the experiment, we set $R=0.5m$, the grid size of the map is $1m \ \times \ 1m$, the moving speed of the robot is $v_r=1 m/s$, the update time step is $T=5s$, and the prediction horizon is $T_D=30s$. We set up 17 different tasks, including 6 delivery mode tasks and 11 pickup mode tasks, randomly distributed in blue squares on the map. To reflect the unpredictable nature of the environment, we set tasks 3-6, 16-17 to be agnostic ahead of time, only to be released at fixed points in time. There are three robots, and one human with a limited load capacity of 60 whose initial positions are fixed. They complete 17 tasks together in the warehouse (see Table \ref{tab3} for details of task settings). For various parameters and weights in the formula, we set $\gamma=0.5$ and $\alpha=0.4$ . We focus on the TA results and the total time required to complete all tasks.


\begin{table*}[]

\centering
\caption{List of simulation initial conditions}
\label{tab3}
\resizebox{\textwidth}{!}{
\begin{tabular}{c|c|c|c|cc|c|c}
\hline
\multicolumn{3}{l|}{The number of robots}  & 3   &\multicolumn{2}{|c|}{The number of human workers}     & 1                                              \\ \hline
\multicolumn{3}{l|}{The maximum load of the robot}                           & 60   & \multicolumn{2}{|c|}{The maximum load of the human}       & 60         \\\hline\hline
Tasks & Mode & \begin{tabular}[c]{@{}c@{}}Time window \\ (min)\end{tabular} & Load & \begin{tabular}[c]{@{}c@{}}Be done \\ by human workers\end{tabular} & \begin{tabular}[c]{@{}c@{}}Be done \\ by robots\end{tabular} & starting point  & ending point \\\hline
1     & delivery    & (0,30)                   & 20   & 0                    & 0                     & (3,5)  & (5,8)  \\\hline
2     & delivery    & (0,30)                  & 20   & 0                           & 0                  & (3,9)  & (3,6)  \\\hline
3     & delivery    & (15,60)                & 20   & 0            & 0                         & (13,5) & (11,11) \\\hline
4          & delivery    & (15,60)                               & 20   & 1                & 0             & (13,9) & (13,11) \\\hline
5          & delivery    & (30,120)           & 20   & 0                & 0            & (7,8)  & (9,4) \\\hline
6     & delivery    & (30,120)          & 20   & 0                            & 0                 & (9,5)  & (3,3)  \\\hline
7     & pickup    & (0,60)                                                      & 20   & 0                                                            & 0                                                            & (3,4)  & (16,7)  \\\hline
8     & pickup    & (0,60)                                                       & 20   & 0                                                            & 0                                                            & (3,10)  & (16,7)  \\\hline
9     & pickup    & (0,60)                                                      & 20   & 0                                                            & 0                                                            & (13,4)  & (16,7)  \\\hline
10     & pickup    & (0,60)                                                      & 20   & 0                                                            & 0                                                            & (13,10)  & (16,7)  \\\hline
11     & pickup    & (0,90)                                                      & 20   & 0                                                            & 0                                                            & (5,5)  & (16,7)  \\\hline
12     & pickup    & (0,90)                                               & 20   & 0                                                            & 0                                                            & (5,10)  & (16,7)  \\\hline
13     & pickup    & (0,90)                                                    & 20   & 0                                                            & 0                                                            & (11,4)  & (16,7)  \\\hline
14     & pickup    & (0,90)                & 20   & 0                                                            & 0                                                            & (11,10)  & (16,7)  \\\hline
15              & pickup    & (0,90)                                                    & 20   & 0                                                            & 0            & (11,8) &(16,7)   \\\hline
16     & pickup    & (30,120)                                                      & 20   & 0                                                            & 0                                                            & (7,4)  & (16,7)  \\\hline
17     & pickup    & (30,120)                                                      & 20   & 0                                                            & 0                                                            & (9,3)  & (16,7)  \\\hline
\end{tabular}}
\end{table*}

\begin{table*}[h]\LARGE
\centering
\caption{Task allocation results for each update time step}
\label{tab4}
\resizebox{\textwidth}{!}{

\begin{tabular}{c|c|c|c|c|c|c|c|c|c|c|c|c|c|c} \hline
 &  & \multicolumn{3}{|c|}{Robot 1} & \multicolumn{3}{|c|}{Robot 2} & \multicolumn{3}{|c|}{Robot 3} & \multicolumn{3}{|c|}{Human 4} &  \\\hline
Time & Unpredictable events & \begin{tabular}[c]{@{}c@{}}Residual \\ load capacity\end{tabular} & \begin{tabular}[c]{@{}c@{}}Current \\ position\end{tabular} & \begin{tabular}[c]{@{}c@{}}Task \\ sequence\end{tabular} & \begin{tabular}[c]{@{}c@{}}Residual \\ load capacity\end{tabular} & \begin{tabular}[c]{@{}c@{}}Current \\ position\end{tabular} & \begin{tabular}[c]{@{}c@{}}Task \\ sequence\end{tabular} & \begin{tabular}[c]{@{}c@{}}Residual \\ load capacity\end{tabular} & \begin{tabular}[c]{@{}c@{}}Current \\ position\end{tabular} & \begin{tabular}[c]{@{}c@{}}Task \\ sequence\end{tabular} & \begin{tabular}[c]{@{}c@{}}Residual \\ load capacity\end{tabular} & \begin{tabular}[c]{@{}c@{}}Current \\ position\end{tabular} & \begin{tabular}[c]{@{}c@{}}Task \\ sequence\end{tabular} & \begin{tabular}[c]{@{}c@{}}Completed \\ tasks\end{tabular} \\\hline
$0$ &  & 60 & (4,2) & 1 & 60 & (12,2) & 13-15-9 & 60 & (12,12) & 14-12-10 & 60 & (4,12) & 8-7-11 & 8,13,14  \\\hline
$T$ &  & 40 & (2,5) & 1 & 40 & (11,4) & 9-10 & 40 & (11,10) & 2 & 40 & (3,10) & 7-11 &   \\\hline
$2T$ &  & 40 & (3,7) & 1 & 40 & (12,2) & 9-10 & 40 & (8,12) & 2 & 40 & (2,6) & 7-12 & 1,7,9  \\\hline
$3T$ & \begin{tabular}[c]{@{}c@{}}The task 3 which must \\ be completed by human, \\ and task 4 publish.\end{tabular} & 60 & (5,8) & 11-12-10 & 20 & (13,4) & 3 & 40 & (3,12) & 2 & 20 & (2,5) & 4 & 11  \\\hline
$4T$ &  & 40 & (5,5) & 12-10 & 0 & (14,6) & 3 & 20 & (3,9) & 2 & 20 & (5,7) & 4 & 2   \\\hline
$5T$ &  & 40 & (6,9) & 12-10 & 0 & (14,11) & 3 & 40 & (3,6) & 15 & 20 & (10,7) & 4 & 12  \\\hline
$6T$ & \begin{tabular}[c]{@{}c@{}}The tasks 5, 6, \\ 16 and 17 publish\end{tabular} & 20 & (6,12) & 5 & 0 & (10,12) & 3 & 40 & (5,7) & 16,17 & 20 & (14,8) & 4 & 3,16 \\\hline
$7T$ &  & 0 & (7,8) & 5 & 20 & (11,11) & 10 & 20 & (7,4) & 6 & 20 & (14,11) & 4 & 4  \\\hline
$8T$ &  & 0 & (9,7) & 5 & 20 & (13,12) & 10 & 20 & (8,2) & 6 & 40 & (13,11) & 17-15 & 5,10  \\\hline
$9T$ & \begin{tabular}[c]{@{}c@{}}All tasks \\ have been allocated\end{tabular} & 20 & (9,4) & 17 & 0 & (14,10) &  & 20 & (10,5) & 6 & 40 & (14,7) & 15 & 17  \\\hline
$10T$ &  & 0 & (10,4) &  & 60 & (16,7) &  & 0 & (10,2) & 6 & 40 & (10,8) & 15 & 15 \\\hline
$11T$ &  & 0 & (12,7) &  &  &  &  & 0 & (5,2) & 6 & 20 & (11,7) &  & 6  \\\hline
$12T$ &  & 60 & (16,7) &  &  &  &  & 0 & (3,3) &  & 60 & (16,7) &  &   \\\hline
$16T$ &  &  &  &  &  &  &  & 60 & (16,7) &  &  &  &  & Done  \\\hline
\end{tabular}}
\end{table*}

\begin{figure}[thpb]  
  \centering 
 \includegraphics[scale=0.6]{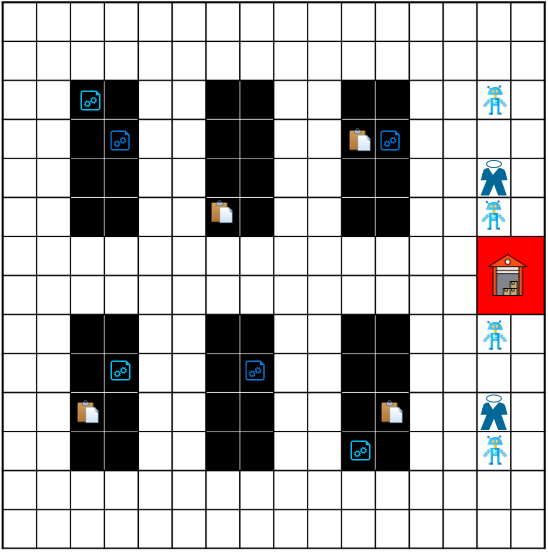} 
  \caption{Path finding results at the third update time step $3T$.}
  \label{fig5} 
\end{figure} 

The \textbf{Problem (HS)$_2$PD} is programmed in MATLAB, and the structure is shown in Algorithm \ref{HSHSPD}. the Yalmip toolbox is used to model and normalize the MIP problems, and the solver uses Gurobi 9.1.2. For the two-layer programming in the first-level structure, the problem is solved using a mix of Fmincon and Gurobi. The TA results at each update time step $T$ are detailed in Table \ref{tab4}. It can be seen that in the ninth update time step, all task requests have been responded {\color{blue}to} and executed. The agents spend the rest of the time sending the picking tasks loaded on them to the warehouse. All tasks are fully completed at the 16th update time step. Fig. \ref{fig5} visually shows the ideal travel route planned for the robot at the third update time step.

\subsection{Experiments on different scales}
{\color{blue}We solved the problem using our proposed framework in ten different scenarios. We have made statistics on ten sets of experimental data, and the completion time, calculation time, average delay and throughput are shown in Figure 7. It can be seen from the figure that the algorithm can run stably at different scales, which verifies the effectiveness of the proposed framework.}
\subsection{Comparison of different solvers}
{\color{blue}For the same experimental scene, we used MATLAB+ Gurobi, MATLAB+GA and MATLAB+Greey solvers to solve the integer programming problem we proposed. Completion time and calculation time are shown in Figure 8. It can be seen from the figure that the integer programming solver can obtain the optimal solution with less computation time, which is due to the problem being modeled as a convex optimization problem that can be solved quickly. Moreover, the robustness of our algorithm in dynamic environment is verified.}
\section{Conclusion} \label{sec5}
This paper has proposed a safe and efficient solution framework for \textbf{Problem (HS)$_2$PD}. The tasks {\color{blue}are allowed to exist in a variety of modes with different requirements.} The acceptance moment of the tasks {\color{blue}is} limited within the time windows. Meanwhile, the framework allows each agent can carry multiple packages. The use of prediction horizon strategy can adapt to unpredictable events, such as newly emerging tasks, abruptly canceled tasks, and the counterproductive behavior of human workers. Further research will consider dynamic obstacle avoidance in maps to improve the working efficiency of (HS)$_2$. Furthermore, reinforcement learning \cite{mazouchi2021conflict} will be used in the future to improve the intelligence and efficiency of (HS)$_2$.

 \end{document}